\documentclass[a4paper,11pt]{article}

\usepackage{pos}
\usepackage{lipsum} %package to generate placeholder text in the following
\usepackage{float}
\usepackage{setspace}
\usepackage{makecell}

\title{NUTRIG: Development of a Novel Radio Self-Trigger for GRAND}

\ShortTitle{NUTRIG: Development of a Novel Radio Self-Trigger for GRAND}

\author*[a]{Pablo Correa}
\author[b]{Jelena Köhler}
\onbehalf{for the GRAND Collaboration{\normalsize \\ \normalfont(a complete list of authors can be found at the end of the proceedings)}\\}

\affiliation[a]{Sorbonne Université, Université Paris Diderot, Sorbonne Paris Cité, CNRS, Laboratoire de Physique Nucléaire et de Hautes Energies
(LPNHE)}
\affiliation[b]{Institute for Astroparticle Physics, Karlsruhe Institute of Technology}

\emailAdd{pablo.correa@lpnhe.in2p3.fr}
\abstract{

The NUTRIG project is dedicated to the development of advanced radio self-trigger methods for large-scale arrays such as the Giant Radio Array for Neutrino Detection (GRAND). The developed techniques are based on features of the radio emission of air showers to perform an efficient online rejection of background. We first describe a first-level trigger (FLT) template-matching method that uses the shape of transient radio pulses measured at the detection-unit level to target those induced by air showers. We present trigger efficiencies and throughput tests of the template-matching FLT in controlled laboratory conditions. Next, we describe the second-level trigger (SLT), which utilizes the measured FLT times and corresponding voltage amplitudes to construct a trigger at the array level. We present offline performances of the SLT, which performs a coarse reconstruction of timing, direction, and polarization of the air shower.
\vspace{4mm}

}

\ConferenceLogo{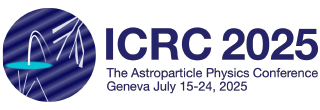}

\FullConference{39th International Cosmic Ray Conference (ICRC2025)\\  15--24 July, 2025\\ Geneva, Switzerland}

\begin{document}

\maketitle

\section{Introduction}
\label{sec:introduction}

The primary goal of the Giant Radio Array for Neutrino Detection (GRAND) \cite{Alvarez_2020,Martineau_2025} is to observe ultra-high-energy astroparticles using radio self-trigger techniques. GRAND is currently in its prototyping phase (see \cite{Ma_2025,deErrico_2025} for details); its main pathfinder array is GRANDProto300 (GP300) located in the Chinese Gobi Desert. As of today, 65 detection units (DUs) have been commissioned at GP300. However, in its next stages (GRAND10k and GRAND200k), GRAND will consist of subarrays with 5,000--10,000 DUs each, with a sparse DU spacing of 1 km. This scale demands an autonomous, efficient, low-cost radio trigger system with minimal inter-DU data transfer.

NUTRIG is the first step towards such a state-of-the-art radio self-trigger. Apart from modeling efforts of air-shower emission between 50--200 MHz \cite{Gulzow_2025,Guelfand_2025}, the frequency band in which GRAND operates, NUTRIG consists of two major parts:

\begin{itemize}
    \item \textbf{First-Level Trigger (FLT) at the DU level.} The classical, so-far used DU-level trigger \cite{Ma_2025} applies a double-threshold technique designed for a nominal rate of 1 kHz at GP300. With NUTRIG, we aim to reduce the DU-level trigger rate to 100 Hz. To achieve this, we developed a template-matching method, described in this work, and an algorithm using a convolutional neural network, which we shall omit here---see \cite{LeCoz_2023,Correa_2024} for more details.
    
    \item \textbf{Second-Level Trigger (SLT) at the array level.} For GRAND prototypes, multiple DUs are required to have an FLT within a coincidence time window to result in a so-called central trigger \cite{Ma_2025}. The expected nominal rate of this array-level trigger is 10 Hz. With NUTRIG, our goal is to reduce this rate to 1 Hz by applying fast online reconstructions on FLT data.
\end{itemize}

These proceedings present the latest results of the NUTRIG FLT and SLT methods, first outlined in \cite{Correa_2024,Kohler_2024}. In Section \ref{sec:database}, we describe novel NUTRIG-dedicated databases, consisting of recent GP300 data and up-to-date air-shower simulations. Next, in Section \ref{sec:flt}, we present the template-matching FLT, including both its offline and online performances. Section \ref{sec:slt} is reserved for the SLT method and its offline performance. Finally, concluding remarks are given in Section \ref{sec:conclusion}.

\section{Databases for NUTRIG Studies}
\label{sec:database}

As summarized in Table \ref{tab:database} and detailed below, we construct several databases for our FLT and SLT studies, using both experimental data and simulations. For the FLT we select individual pulses, while for the SLT we select array-level events.

\begin{table}[t!]
\small
\centering
\begin{tabular}{lccccc}
\Xhline{2\arrayrulewidth}
Database name & Data type & Number of & SNR & FIR filter & FLT-0 setting  \\
 & & pulses/events & & [offline] & [offline]   \\
\Xhline{1\arrayrulewidth}
FLT background & MD pulses & 6,628 & 4--8 & Yes & FLT-0-light \\
FLT signal & Sim.~pulses + stat.~MD & 8,000 & 4--8 & Yes & FLT-0-light \\
\Xhline{1\arrayrulewidth}
SLT background & CD events & 2,146 ($\geq$5 DUs) & $>$5 & Yes & FLT-0-optimal \\
SLT signal & Sim.~events + stat.~MD & 5,620 ($\geq$5 DUs) & $>$5 & Yes & FLT-0-optimal \\
\Xhline{1\arrayrulewidth}
Electric transformer & CD pulses& 1,368 & $>$5 & No & --- \\
Cosmic-ray candidates & CD pulses & 225 & $>$5 & No & --- \\
\Xhline{2\arrayrulewidth}
\end{tabular}
\caption{Summary of the different databases used in this work, which consist of experimental GP300 data in the form of MD and CD, as well as air-shower simulations (sim.) superimposed to stationary (stat.) MD. See text for more details.}
\label{tab:database}
\end{table}

\subsection{Selection of Pulses from GP300 Data}
\label{sec:database_background}

As extensively described in \cite{AlvesBatista_2024}, a DU of GP300 consists, among other components, of a three-arm butterfly antenna---two dipoles along the North-South ($X$) and East-West ($Y$) axes, and a monopole along the vertical ($Z$) axis---and its front-end electronics board (FEB). The FEB contains a 14-bit analog-to-digital converter (ADC) with a sampling rate of 500 Msps as well as a Xilinx system on chip (SoC). The SoC consists of a field-programmable gate array (FPGA) that performs the GP300 double-threshold trigger, which we will name FLT-0 in this work, and a central processing unit (CPU) for event building. The NUTRIG FLT component, which we label as FLT-1, is implemented on the CPU (see Section \ref{sec:flt_online}).

At GP300, the observed radio-frequency interference (RFI) is dominated by airplanes and the electric transformer of a mine located 15 km from the site \cite{Lavoisier_2025}. To mitigate the high-frequency noise of this electric transformer as well as aeronautic communication lines, a finite impulse response (FIR) lowpass filter with a cutoff frequency of 115 MHz is applied by the FPGA before evaluating the FLT-0. For triggered events, 2048 ns of the unfiltered ADC traces are recorded, for each polarization, by the central data-acquisition (DAQ) system.

In this work, we first construct the FLT background database by identifying pulses embedded in forced-trigger monitoring data (MD) taken in February 2025, with a total runtime of $\sim$12 hours at a rate of 20 Hz. We mimic the online conditions by passing the data through the FIR filter and FLT-0 algorithm, which can trigger on the $X$ or $Y$ channels. Here, we use relaxed conditions on the FLT-0 parameters (FLT-0-light) that allow us to select pulses down to relatively low signal-to-noise ratios $\rm SNR \gtrsim 4$. We define the SNR as maximum of the pulse divided by the root-mean-square (RMS) of the trace excluding the pulse. From the background pulses that pass the FLT-0-light, we then randomly select 2,000 traces for both $\rm 4 \leq SNR < 5$ and $\rm 5 \leq SNR < 6$. Due to the limited MD runtime, only 355 and 74 traces are selected for $\rm 6 \leq SNR < 7$ and $\rm 7 \leq SNR < 8$, respectively.

Next, we construct the SLT background database, which consists of coincidence data (CD; runtime of $\sim$30 hours) taken in April--May 2025. The CD contains events that passed the online GP300 central trigger, for which at least 4 DUs are required to yield an FLT-0 (which was exlusively applied to the $Y$ channel) within a time window of 11 $\mu$s. For consistency with the SLT signal database (Section \ref{sec:database_signal}), we reapply the FIR filter and FLT-0 algorithm to the recorded data. In this case, we use FLT-0 settings that are optimized for a maximal signal selection efficiency while keeping the expected mean trigger rate below 100 Hz in the aforementioned MD (FLT-0-optimal). We select 2,146 background events where at least 5 coincident DUs pass the FLT-0-optimal with an $\rm SNR > 5$.

Finally, for an FLT-1 case study presented in Section \ref{sec:flt_offline}, we make a selection of 1,368 pulses originating from the electric transformer. These are randomly selected from a CD run of March 2025 in which no online FIR filter was applied. In addition, we select the 225 pulses of the cosmic-ray candidate CD events identified in \cite{Lavoisier_2025}.

\subsection{Selection of Pulses from Air-Shower Simulations}
\label{sec:database_signal}

We start from 25,000 ZHAireS \cite{Alvarez-Muniz_2011} simulations---half proton primaries, half iron primaries---with energies, zenith angles, and azimuth angles covering $\log_{10} (E/\mathrm{eV}) \in [16.5,18.6]$ uniformly, $\theta/{\rm deg} \in [30.6,87.3]$ following $\log_{10}(1 / \cos \theta)$, and $\phi/{\rm deg} \in [0,360]$ uniformly, respectively. Electric fields are simulated for all DU positions of a preliminary GP300 layout (including a denser infill array), and are then propagated through the antenna response and radio-frequency (RF) chain up to the ADC using GRANDlib \cite{AlvesBatista_2024}. Note that the simulated RF chain is in good agreement with GP300 data for the $X$ and $Y$ polarizations.

The simulated air-shower ADC pulses are subsequently superimposed to stationary-background traces. This stationary background is selected from $\sim$3 hours of GP300 MD data taken at a rate of 20 Hz in February 2025. We limit ourselves to data of 6 stable DUs that yield nominal power spectrum densities (see \cite{Ma_2025} for an example), and we also correct baselines that are offset from zero. To be classified as stationary, traces are required to have a $\rm RMS \leq 20$ ADC counts, and a maximum amplitude below $\rm 5 \times RMS$, for all three polarizations.

From the above simulations, we construct both the FLT and SLT signal databases using the exact same procedures as for the FLT and SLT background databases. For the FLT signal database, we apply the FIR filter and FLT-0-light configuration, selecting 2,000 random pulses per SNR bin in [4,5,...,8]. On the other hand, for the SLT signal database, we apply the FIR filter and FLT-0-optimal settings, yielding 5,620 air-shower events with at least 5 coincident DUs with $\rm SNR > 5$.

\section{First-Level Trigger}
\label{sec:flt}

\subsection{Template-Matching Algorithm}
\label{sec:template_algo}

At the ADC level, the shape of an air-shower signal is dominated by the RF-chain response. Hence, we can restrict ourselves to a small number of ADC templates to describe these air-shower signals. Motivated by the time constraints of online processing (see Section \ref{sec:flt_online}), we select 5 ADC templates from the air-shower simulations described in Section \ref{sec:database_signal}, after applying the FIR filter. These air-shower ADC templates are limited to a length of 200 ns. Since the electric-field shape of an air-shower pulse depends on the opening angle to the shower axis, $\omega$, we select the most representative ADC template in 5 bins between $0 \leq \omega/\omega_c \leq 2$, with $\omega_c$ the Cherenkov angle. The most representative template per bin is the one that yields the maximum average cross-correlation (see below) with the other templates in that bin. Note that we use the same 5 templates for both $X$ and $Y$ polarizations, since they have identical RF chains. 

Inspired by \cite{Barwick_2016}, for each polarization\footnote{With a corresponding selection of templates, this method can directly be expanded to include the $Z$ polarization.} $i \in \lbrace X,Y \rbrace$ and template index $j \in \lbrace 1,2,...,5 \rbrace$, we compute the cross-correlation $\rho_{ij} \in [-1,1]$ of an input trace $V_i$ with a template $T_{ij}$---both normalized over the sliding 200-ns integration window---as
\begin{equation}
    \rho_{ij}(\tau) = \int T_{ij}(t)\, V_i(t+\tau)\, \mathrm{d}t,
    \label{eq:corr}
\end{equation}
where $\tau \in [t_{\text{FLT-0}} - 20~\mathrm{ns}, t_{\text{FLT-0}} + 20~\mathrm{ns}]$. The cross-correlation serves as a measure of the match between the input trace and the template (see also Fig.~\ref{fig:flt_mine}). Therefore, we determine the best-match cross-correlation as $\rho \equiv \max_{i,j,\tau} | \rho_{ij}(\tau) | \in [0,1]$, where we take the absolute value since $\rho_{ij} < 0$ simply indicates an opposite polarity between the trace and the template. This best-match value $\rho$ is the main discrimination parameter for the template FLT-1. Note that our method has been simplified w.r.t.~our previous work \cite{Correa_2024} in order to increase computing speed without loss of performance.

\subsection{Offline Analysis Results}
\label{sec:flt_offline}

\begin{figure}[t]
    \centering
    \includegraphics[width=0.490\textwidth]{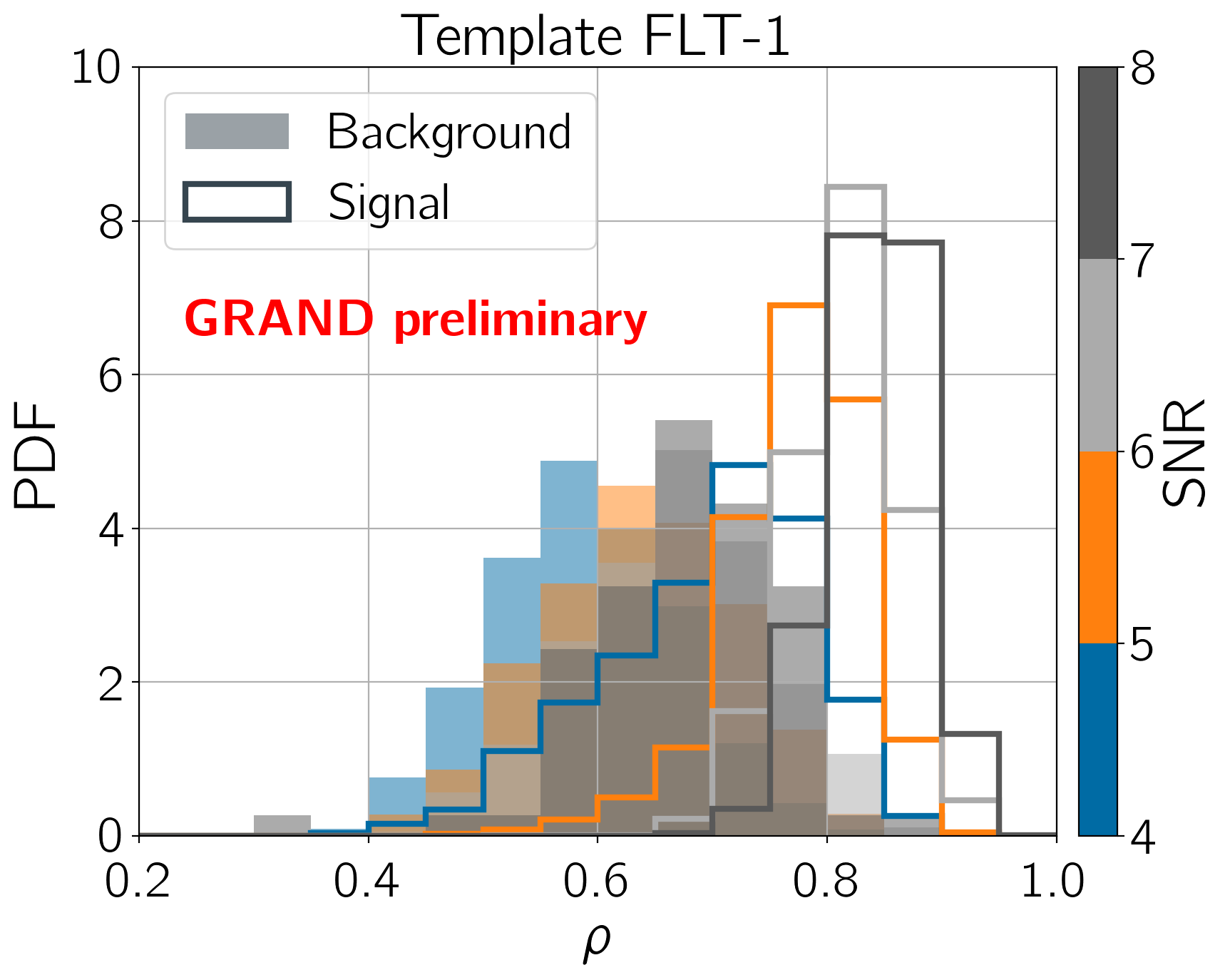}
    \includegraphics[width=0.500\textwidth]{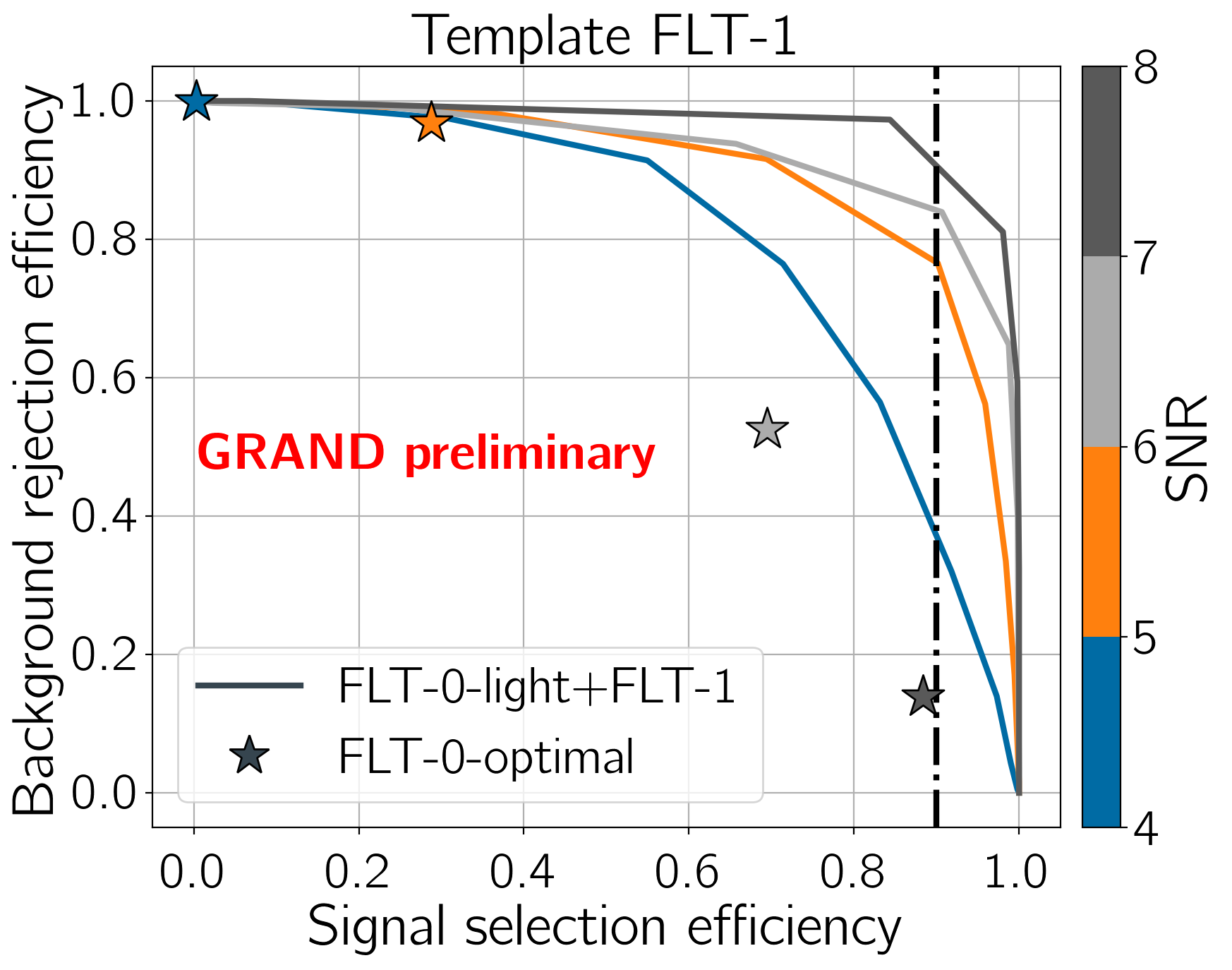}
    \caption{\textit{Left}: Distribution of the best-match cross-correlation value $\rho$ of the template-matching FLT-1 method for both the FLT background database (filled histograms) and FLT signal database (empty histograms). The different colors represent the results for different SNR bins. \textit{Right}: Signal selection efficiencies versus background rejection efficiencies as a function of SNR. These are shown for the FLT-0-light+FLT-1 configuration (filled lines), corresponding to the left panel, as well as the FLT-0-optimal configuration (stars). The dash-dotted line represents a signal-selection efficiency of 90\%.}
    \label{fig:flt_efficiencies}
\end{figure}

We apply our template-matching FLT-1 algorithm to the FLT background and signal databases described in Section \ref{sec:database}. The left panel of Fig.~\ref{fig:flt_efficiencies} shows the distributions of the best-fit cross-correlation values $\rho$ for both background and signal, each in different SNR bins. From these distributions, we use a sliding threshold value $\rho_{\rm thresh}$ to obtain the signal selection (fraction with $\rho \geq \rho_{\rm thresh}$) and background rejection (fraction with $\rho < \rho_{\rm thresh}$) efficiencies, shown in the right panel of Fig.~\ref{fig:flt_efficiencies}. We find that with increasing SNR, the template-matching FLT-1 yields a more significant separation of background-RFI pulses from air-shower signal pulses. In particular, for $\rm SNR > 5$, we find that if we impose a signal selection efficiency of 90\%, we can reject $\gtrsim$75\% of the background pulses. %Note that compared to our previous work \cite{Correa_2024}, the results shown here are more representative of the offline FLT-1 performance, since now we are also binning the background distribution in SNR.

We compare the above performance of the FLT-0-light+FLT-1 logic to the FLT-0-optimal configuration without any template-matching FLT-1 component. To do so, we apply the FLT-0-optimal algorithm directly to the MD data used for the FLT background database (Section \ref{sec:database_background}), and compute selection (resp.~rejection) efficiencies per SNR bin as the fraction of traces that are accepted (resp.~rejected). These are indicated by the stars in the right panel of Fig.~\ref{fig:flt_efficiencies}. With increasing SNR, we find that for the same signal selection efficiency achieved by the FLT-0-optimal---which becomes sensitive for 
$\rm SNR \gtrsim 6$---the FLT-0-light+FLT-1 improves significantly on the background rejection (e.g.~$\sim$90\% compared to $\lesssim$20\% for $\rm 7 \leq SNR < 8$).

\begin{figure}
    \centering
    \includegraphics[width=0.483\textwidth]{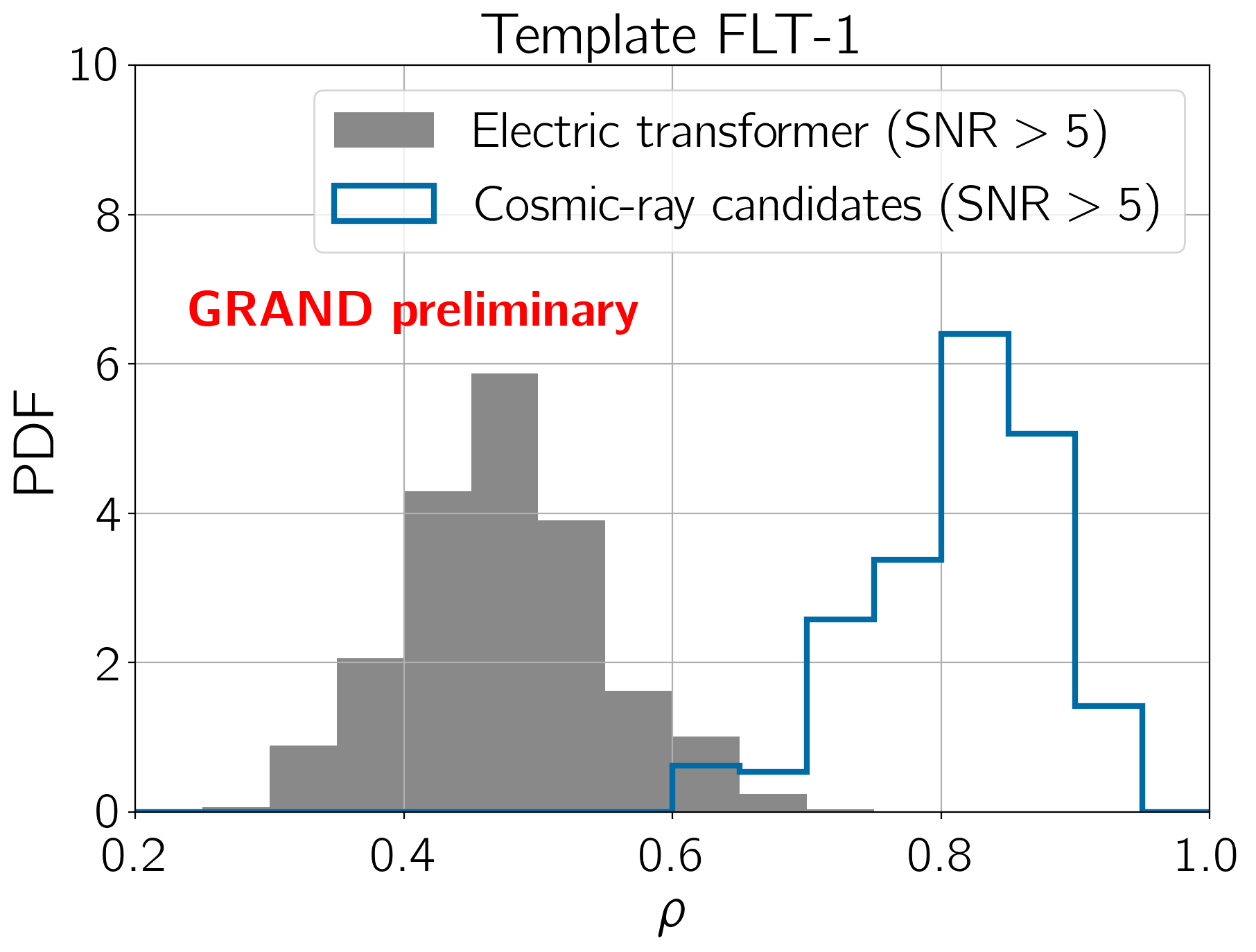}
    \includegraphics[width=0.507\textwidth]{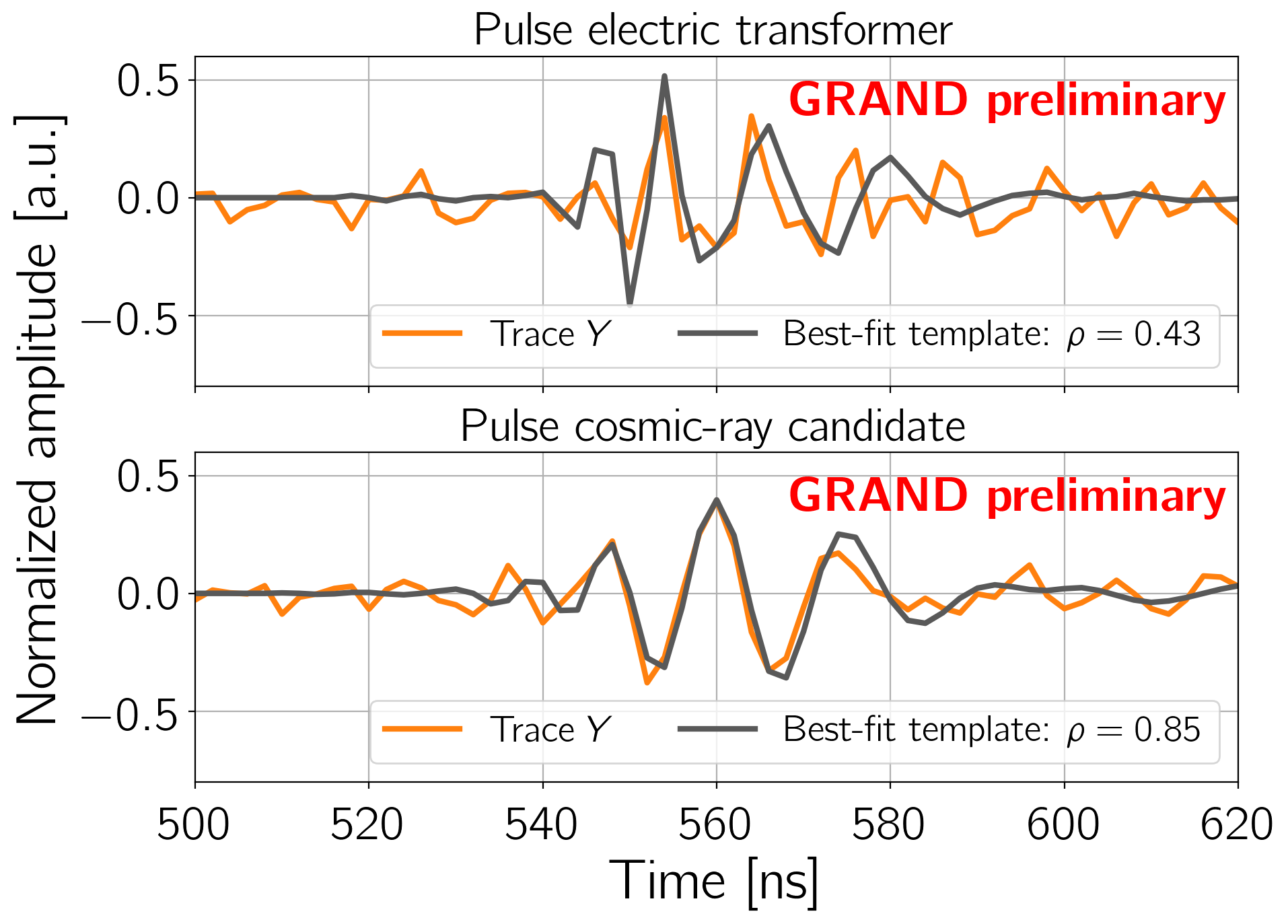}
    \caption{\textit{Left}: Distribution of the best-fit cross-correlation value $\rho$ of the template-matching FLT-1 method for background-RFI pulses of the electric transformer (filled histogram) and the pulses of the GP300 cosmic-ray candidates (empty histogram). \textit{Right}: Examples of the template fit to a background-RFI pulse (top) and cosmic-ray candidate pulse (bottom). Note that no FIR filtering is applied to pulses in this treatment.}
    \label{fig:flt_mine}
\end{figure}

Finally, to illustrate the power of the template FLT-1 method, we apply it to both the electric transformer database and the cosmic-ray candidate database described in Section \ref{sec:database_background}. In this case study, no FIR filter is applied to choose the 5 optimal templates (Section \ref{sec:template_algo}). We feed the data directly to the template-matching FLT-1 without passing through the FLT-0-light. The resulting $\rho$-distributions are shown in the left panel of Fig.~\ref{fig:flt_mine}. We find that the template-matching FLT-1 method yields an excellent separation of the electric-transformer pulses and the cosmic-ray candidate pulses. This is a consequence of the fact that the RFI pulses of the electric transformer have a dominant high-frequency component, as shown in the right panel of Fig.~\ref{fig:flt_mine}.

\subsection{Online Analysis Results}
\label{sec:flt_online}

Online tests of the FLT-1 have been performed at the LPNHE test bench, described in \cite{Correa_2023}. We successfully ported the template-matching FLT-1 algorithm to the DU-DAQ software (written in \texttt{C++}) on the CPU of the Xilinx SoC on the FEB. We performed an isolated test of the algorithm on this CPU, and found that it can treat signals with rates up to $\sim$5.5 kHz using 5 templates. This is well above the nominal 1 kHz that is expected to be fed to the FLT-1 by the FLT-0. In addition, for a test trace fed to the FEB using an arbitrary-wave generator, we successfully reproduced the same best-fit correlation values in the online and offline template-matching FLT-1 treatments. In the near future, our goal is to perform stability tests in realistic conditions at GRAND@Nançay \cite{Correa_2023}, with the aim to deploy the template-matching FLT-1 at GP300 by Fall 2025.

\section{Second-Level Trigger}
\label{sec:slt}
The SLT algorithm, which is still under development, currently consists of three main steps. In these steps, we perform selection cuts on reconstructions of the arrival time, arrival direction (zenith and azimuth angles), and polarization of the air-shower emission. In particular, we aim to maximize the selection (resp.~rejection) efficiency of events in the SLT signal database (resp.~SLT background database), as described below.

\subsection{Step 1: Timing Outlier Rejection}\label{sec:SLT_timing}

To exclude individual DUs that were not triggered by an air shower, we first estimate the air-shower arrival direction using the pulse arrival times provided by the FLT. For that, we apply an analytic plane wave front (PWF) reconstruction method \cite{arsene}. We then apply a cut on the time difference $\Delta t$ between the measured arrival times and the PWF expectation, $\Delta t/{\rm ns} \in [-20,20]$.

\subsection{Step 2: Consistency of Arrival Times and Signal Strength}\label{sec:slt_conic}

In the second step of the SLT algorithm, we use the conic method presented in a previous work \cite{Kohler_2024} to model the signal-strength distribution of an air shower as a cone that intersects the ground, resulting in an elliptical footprint. The shower zenith angle $\theta$ and azimuth angle $\varphi$ are determined by their relation to the semi-major and semi-minor axes of an ellipse fitted to the signal distribution in the air-shower footprint. They are then compared to the zenith and azimuth angles reconstructed with the PWF method. Note that the conic method yields a degeneracy for $\varphi/{\rm deg} \in [0,180]$ and $\varphi/{\rm deg} \in [180,360]$, which is taken into account in the comparison.

\begin{figure}
    \centering
    \includegraphics[width=0.63\linewidth]{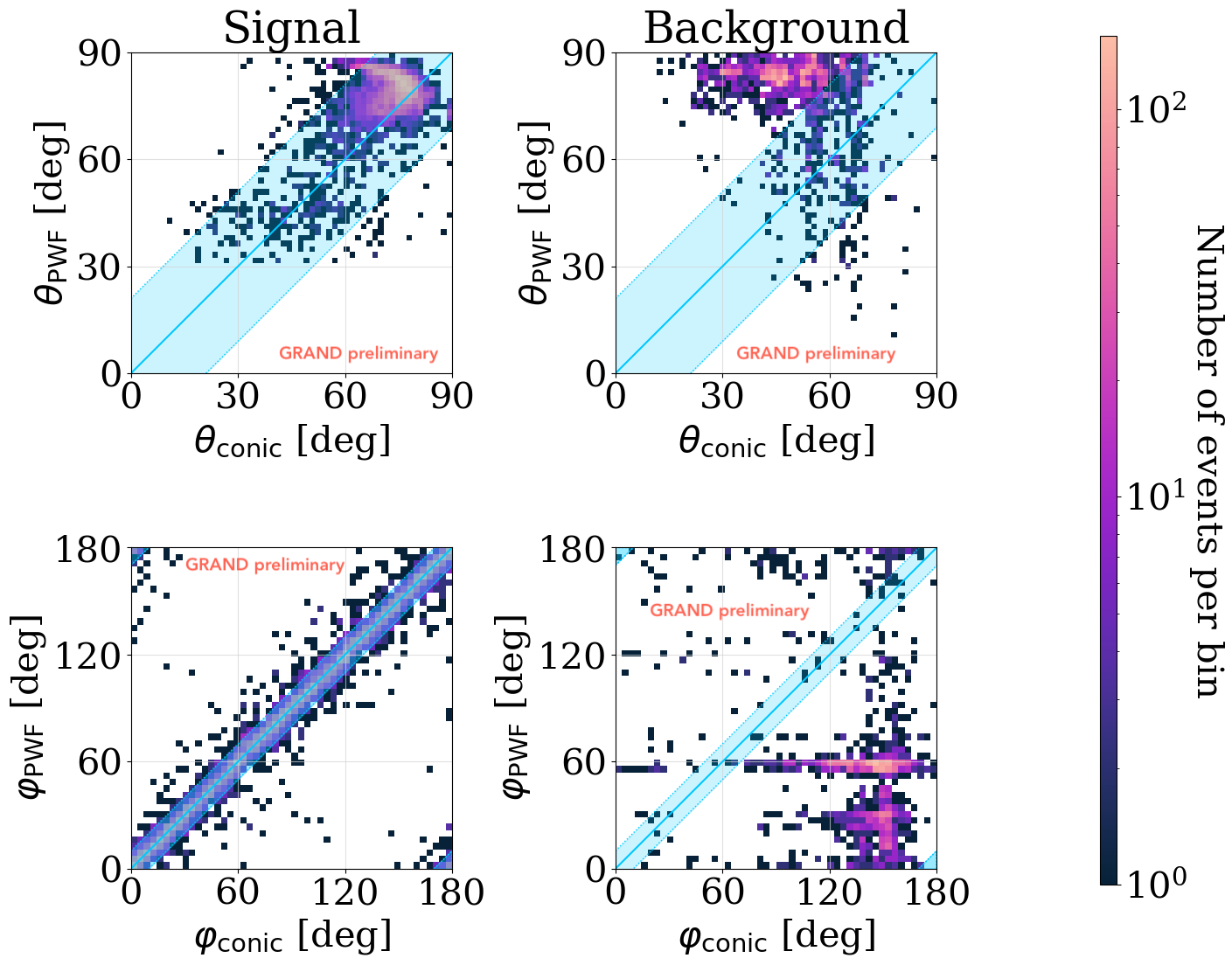}
    \caption{Comparison of the reconstructed zenith angles (top row) and azimuth angles (bottom row; degeneracy for $\varphi_{\rm conic}/{\rm deg} \in [180,360]$ taken into account) obtained with the PWF and conic methods, for both the SLT signal database (left column) and SLT background database (right column). In each plot, the dashed cyan line represents an ideal one-to-one correlation, while the cyan band indicates the cut values applied by the SLT.}
    \label{fig:slt_conic}
\end{figure}

Figure~\ref{fig:slt_conic} shows the zenith and azimuth angle distributions obtained with both methods for the SLT signal and background databases. The PWF reconstructions of signal events match the underlying zenith and azimuth angle distributions of the simulations (Section \ref{sec:database_signal}), while the PWF reconstructions of the background events are consistent with a dominant contribution of noise pulses received from airplanes (see \cite{Lavoisier_2025}). The conic method yields a good agreement with the PWF for the azimuth angle reconstruction of both signal and background events. However, due to edge effects at the borders of the array as well as the variable spacing of DUs near the simulated infill area \cite{Kohler_2024}, poorer zenith angle reconstructions are currently achieved with the conic method. These can be improved by performing the fit only to the DUs with a strong signal and by filtering out the infill to create an even grid, and will generally be less of a problem for larger arrays.

For the SLT, we perform cuts on both the difference in zenith angle, $\Delta \theta = \theta_{\rm conic} - \theta_{\rm PWF}$, and azimuth angle, $\Delta \varphi = \varphi_{\rm conic} - \varphi_{\rm PWF}$. We find that we can achieve 95\% signal selection efficiency for $\Delta \theta / {\rm deg} \in [-21,21]$ and $\Delta \varphi / {\rm deg} \in [-10,10]$, yielding background rejection efficiencies of 82\% and 98\%, respectively. We note that the azimuth reconstruction yields the most powerful signal-background separation in the SLT method given the current SLT signal and background databases.

\subsection{Step 3: Polarization}\label{sec:slt_polarization}

A radio signal polarized along the $\mathbf{v} \times \mathbf{B}$ axis (with $\mathbf{v}$ the shower velocity and $\mathbf{B}$ the geomagnetic field) is a strong indicator of an air-shower signal, as the geomagnetic emission represents its dominant emission component. We estimate the polarization of electric-field signals in the ground plane using the method developed in \cite{Chiche_2022}. Here, the polarization is estimated via the ratio $E_b / E_\text{tot}$, with $E_b$ the magnitude of the projection of the electric field along $\mathbf{B}$, and $E_\text{tot}$ the total electric-field amplitude. For trigger applications, this method is analogously suggested for voltages, such that $E$ can be replaced by $V$. However, this means that the antenna response is not deconvolved, which is expected to decrease performance. Indeed, we find that the distribution of $V_b / V_\text{tot}$ is similar for background and signal events of the SLT database. This is also likely resulting from the online FLT-0 conditions of the CD data that were used to construct the SLT background database, namely triggering on pulses in the East-West antenna polarization. Nevertheless, to obtain a signal selection efficiency of 95\%, we perform the cut $V_b / V_\text{tot} < 0.2$, which yields a background rejection efficiency of 23\%. In further studies, this might be improved by adding a simplified reconstruction of the electric field vector from the voltage amplitudes, as long as its execution time is suitable for online SLT conditions.

\section{Conclusion}
\label{sec:conclusion}
In this work we presented the latest developments of the NUTRIG project using dedicated, up-to-date databases of GP300 data and air-shower simulations. First, we developed a template-matching FLT algorithm at the DU level that yields a background rejection efficiency of $\gtrsim$75\% for a fixed signal selection efficiency of $90\%$ for $\rm SNR > 5$. This is a significant improvement compared to the current GP300 FLT algorithm, which only rejects $\lesssim$20\% of the background for $\rm SNR > 7$. Moreover, we found that the template-matching FLT yields a good separation of events from the electric transformer and the GP300 cosmic-ray candidates. In addition, we determined that in online conditions, the isolated template-matching FLT is capable of processing an event rate of $\sim$5.5 kHz.

Second, we developed an SLT algorithm that combines reconstructions of arrival time, direction, and polarization from both timing and signal-strength information at the array level. For events with $\geq$5 DUs with $\rm SNR > 5$, we find that for a required signal selection efficiency of 95\%, the reconstructions of zenith angle, azimuth angle, and polarization yield background rejection efficiencies of 82\%, 98\%, and 23\%, respectively. In particular, the performance of the azimuth-angle reconstruction from the signal-strength footprint is promising for a future, completed SLT implementation.

% Bibtex references:
\begingroup
\setstretch{0.3}
\fontsize{9.5pt}{0pt}\selectfont
\bibliographystyle{ICRC}
\bibliography{references}
\endgroup

% Alternatively, you can include references by hand:
%\begin{thebibliography}{99}
%\bibitem{...}
%
%\end{thebibliography}

\clearpage

%The following list of authors, affiliations and funding agencies will be updated at the day of submission. The following template is a placeholder generated with the member list updated to April 1, 2025..
\section*{Full Author List: GRAND Collaboration}

\scriptsize
\noindent
J.~Álvarez-Muñiz$^{1}$, R.~Alves Batista$^{2, 3}$, A.~Benoit-Lévy$^{4}$, T.~Bister$^{5, 6}$, M.~Bohacova$^{7}$, M.~Bustamante$^{8}$, W.~Carvalho$^{9}$, Y.~Chen$^{10, 11}$, L.~Cheng$^{12}$, S.~Chiche$^{13}$
% ...
\\
\\
$^{1}$Departamento de Física de Particulas \& Instituto Galego de Física de Altas Enerxías, Universidad de Santiago de Compostela, 15782 Santiago de Compostela, Spain \\
$^{2}$Institut d'Astrophysique de Paris, CNRS  UMR 7095, Sorbonne Université, 98 bis bd Arago 75014, Paris, France \\
$^{3}$Sorbonne Université, Université Paris Diderot, Sorbonne Paris Cité, CNRS, Laboratoire de Physique Nucléaire et de Hautes Energies (LPNHE), 4 place Jussieu, F-75252, Paris Cedex 5, France \\
$^{4}$Université Paris-Saclay, CEA, List,  F-91120 Palaiseau, France \\
$^{5}$Institute for Mathematics, Astrophysics and Particle Physics, Radboud Universiteit, Nijmegen, the Netherlands \\
$^{6}$Nikhef, National Institute for Subatomic Physics, Amsterdam, the Netherlands \\
$^{7}$Institute of Physics of the Czech Academy of Sciences, Na Slovance 1999/2, 182 00 Prague 8, Czechia \\
$^{8}$Niels Bohr International Academy, Niels Bohr Institute, University of Copenhagen, 2100 Copenhagen, Denmark \\
$^{9}$Faculty of Physics, University of Warsaw, Pasteura 5, 02-093 Warsaw, Poland \\
$^{10}$Key Laboratory of Dark Matter and Space Astronomy, Purple Mountain Observatory, Chinese Academy of Sciences, 210023 Nanjing, Jiangsu, China \\
$^{11}$School of Astronomy and Space Science, University of Science and Technology of China, 230026 Hefei Anhui, China \\
$^{12}$National Astronomical Observatories, Chinese Academy of Sciences, Beijing 100101, China \\
$^{13}$Inter-University Institute For High Energies (IIHE), Université libre de Bruxelles (ULB), Boulevard du Triomphe 2, 1050 Brussels, Belgium \\
% ...

%%%%%%%%%%%%%%%%%%%%%%%%%%%%%%%%%%%%%%%%%%%%%%%%%%%%%%%%%%%%%%
%%%%%%%%%%%%%%%%%%%%%%%%%%%%%%%%%%%%%%%%%%%%%%%%%%%%%%%%%%%%%%

\subsection*{Acknowledgments}

\noindent
The GRAND Collaboration is grateful to the local government of Dunhuag during site survey and deployment approval, to Tang Yu for his help on-site at the GRANDProto300 site, and to the Pierre Auger Collaboration, in particular, to the staff in Malarg\"ue, for the warm welcome and continuing support.
The GRAND Collaboration acknowledges the support from the following funding agencies and grants.
%%%%
\textbf{Brazil}: Conselho Nacional de Desenvolvimento Cienti\'ifico e Tecnol\'ogico (CNPq); Funda\c{c}ão de Amparo \`a Pesquisa do Estado de Rio de Janeiro (FAPERJ); Coordena\c{c}ão Aperfei\c{c}oamento de Pessoal de N\'ivel Superior (CAPES).
%%%%
\textbf{China}: National Natural Science Foundation (grant no.~12273114); NAOC, National SKA Program of China (grant no.~2020SKA0110200); Project for Young Scientists in Basic Research of Chinese Academy of Sciences (no.~YSBR-061); Program for Innovative Talents and Entrepreneurs in Jiangsu, and High-end Foreign Expert Introduction Program in China (no.~G2023061006L); China Scholarship Council (no.~202306010363); and special funding from Purple Mountain Observatory.
%%%%
\textbf{Denmark}: Villum Fonden (project no.~29388).
%%%%
\textbf{France}: ``Emergences'' Programme of Sorbonne Universit\'e; France-China Particle Physics Laboratory; Programme National des Hautes Energies of INSU; for IAP---Agence Nationale de la Recherche (``APACHE'' ANR-16-CE31-0001, ``NUTRIG'' ANR-21-CE31-0025, ANR-23-CPJ1-0103-01), CNRS Programme IEA Argentine (``ASTRONU'', 303475), CNRS Programme Blanc MITI (``GRAND'' 2023.1 268448), CNRS Programme AMORCE (``GRAND'' 258540); Fulbright-France Programme; IAP+LPNHE---Programme National des Hautes Energies of CNRS/INSU with INP and IN2P3, co-funded by CEA and CNES; IAP+LPNHE+KIT---NuTRIG project, Agence Nationale de la Recherche (ANR-21-CE31-0025); IAP+VUB: PHC TOURNESOL programme 48705Z. 
%%%%
\textbf{Germany}: NuTRIG project, Deutsche Forschungsgemeinschaft (DFG, Projektnummer 490843803); Helmholtz—OCPC Postdoc-Program.
%%%%
\textbf{Poland}: Polish National Agency for Academic Exchange within Polish Returns Program no.~PPN/PPO/2020/1/00024/U/00001,174; National Science Centre Poland for NCN OPUS grant no.~2022/45/B/ST2/0288.
%%%%
\textbf{USA}: U.S. National Science Foundation under Grant No.~2418730.
%%%
Computer simulations were performed using computing resources at the CCIN2P3 Computing Centre (Lyon/Villeurbanne, France), partnership between CNRS/IN2P3 and CEA/DSM/Irfu, and computing resources supported by the Chinese Academy of Sciences.

\end{document}